\documentclass[aps,prl,reprint,superscriptaddress]{revtex4-1}

\usepackage{graphicx}

\begin{document}


\title{Quantum Hall Effect in a Josephson Junction}


\author{Stefano Guiducci}
\affiliation{NEST, Istituto Nanoscienze-CNR and Scuola Normale Superiore, Piazza San Silvestro 12, 56127 Pisa, Italy}

\author{Matteo Carrega}
\affiliation{NEST, Istituto Nanoscienze-CNR and Scuola Normale Superiore, Piazza San Silvestro 12, 56127 Pisa, Italy}

\author{Giorgio Biasiol}
\affiliation{IOM CNR, Laboratorio TASC, Area Science Park, 34149 Trieste, Italy}

\author{Lucia Sorba}
\affiliation{NEST, Istituto Nanoscienze-CNR and Scuola Normale Superiore, Piazza San Silvestro 12, 56127 Pisa, Italy}

\author{Fabio Beltram}
\affiliation{NEST, Istituto Nanoscienze-CNR and Scuola Normale Superiore, Piazza San Silvestro 12, 56127 Pisa, Italy}

\author{Stefan Heun}
\email[]{stefan.heun@nano.cnr.it}
\affiliation{NEST, Istituto Nanoscienze-CNR and Scuola Normale Superiore, Piazza San Silvestro 12, 56127 Pisa, Italy}


\date{\today}

\begin{abstract}
Hybrid superconductor/semiconductor devices constitute a powerful platform where intriguing topological properties can be investigated. Here we present fabrication methods and analysis of Josephson junctions formed by a high-mobility InAs quantum-well bridging two Nb superconducting contacts. We demonstrate  supercurrent flow with  transport measurements, critical temperature of 8.1 K, and critical fields of the order of 3~T. Modulation of supercurrent amplitude can be achieved by acting on two side gates lithographed close to the two-dimensional electron gas. Low-temperature measurements reveal also well-developed quantum Hall plateaus, showing clean quantization of Hall conductance. Here the side gates can be used to manipulate channel width and electron carrier density in the device. These findings demonstrate the potential of these hybrid devices to investigate the coexistence of superconductivity and Quantum Hall effect and constitute the first step in the development of new device architectures hosting topological states of matter.\end{abstract}

\pacs{}

\maketitle

\section{Introduction}

Recently, notions of electronic-band physics combined with typical tools of geometry and topology brought to the prediction of new states of matter \cite{Qi2011,Haldane2017,Hasan2010,Stern2008}, with potential applications in quantum technologies. First experimental evidence of topological states of matter goes back to the 1980s with the discovery of the quantum Hall (QH) effect \cite{Klitzing1980,Laughlin1983,Tsui1982,Saminadayar1997,Bid2010}. Here, the ultra-precise degree of conductance quantization is linked to the topological properties of the QH state. Indeed, owing to their non-trivial topology, metallic edge states are robust against disorder and weak perturbations and lead to the concept of topological protection. Among all applications exploiting this intrinsic robustness, the possibility to store quantum information in a stable fashion has recently played a major role \cite{Haldane2017,Nayak2008}. The recent discovery of new bound states with non-trivial braiding properties, such as Majorana states \cite{Mourik2012,Albrecht2016,Mong2014,Alicea2016}, renewed the interest in the study of hybrid superconductor (SC)/semiconductor systems \cite{Wan2015,Tiira2017,Wickramasinghe2018}. Among all, a promising candidate for hosting Majorana fermions is a QH state in proximity to a SC \cite{Mason2016,Stone2011,Ostaay2011}. Experimental evidence for the coexistence of superconductivity and QH is still scarce, however, different SC materials with large critical field can be envisioned that withstand magnetic fields sufficiently high to induce the QH regime in new hybrid devices. Indeed, very recently signatures of superconducting correlations in QH channels were reported in Josephson junctions (JJ) using graphene and III-V based 2D electron gas (2DEG) \cite{Amet2016,Lee2017,Shalom2016,Wu2017}.

Here, we present results on the fabrication methods and analysis of JJ devices with a III-V high-mobility 2DEG interfaced with Nb superconducting contacts. We report transport measurements, demonstrate supercurrent flow and a critical temperature of $8.1$~K, with critical fields as high as $3$~T. Tuning of Josephson current is achieved by means of additional side gates. Moreover, we show that the same samples support well-developed QH plateaus reaching filling factor $\nu=2$ at $B=3$~T. These findings demonstrate the potential of these hybrid devices to investigate the coexistence of SC and QH. We believe they can represent the first step in the development of new device architectures hosting topological states of matter.

\section{Experimental Details}

\begin{figure}[t]
  \includegraphics[width=\linewidth,keepaspectratio=true]{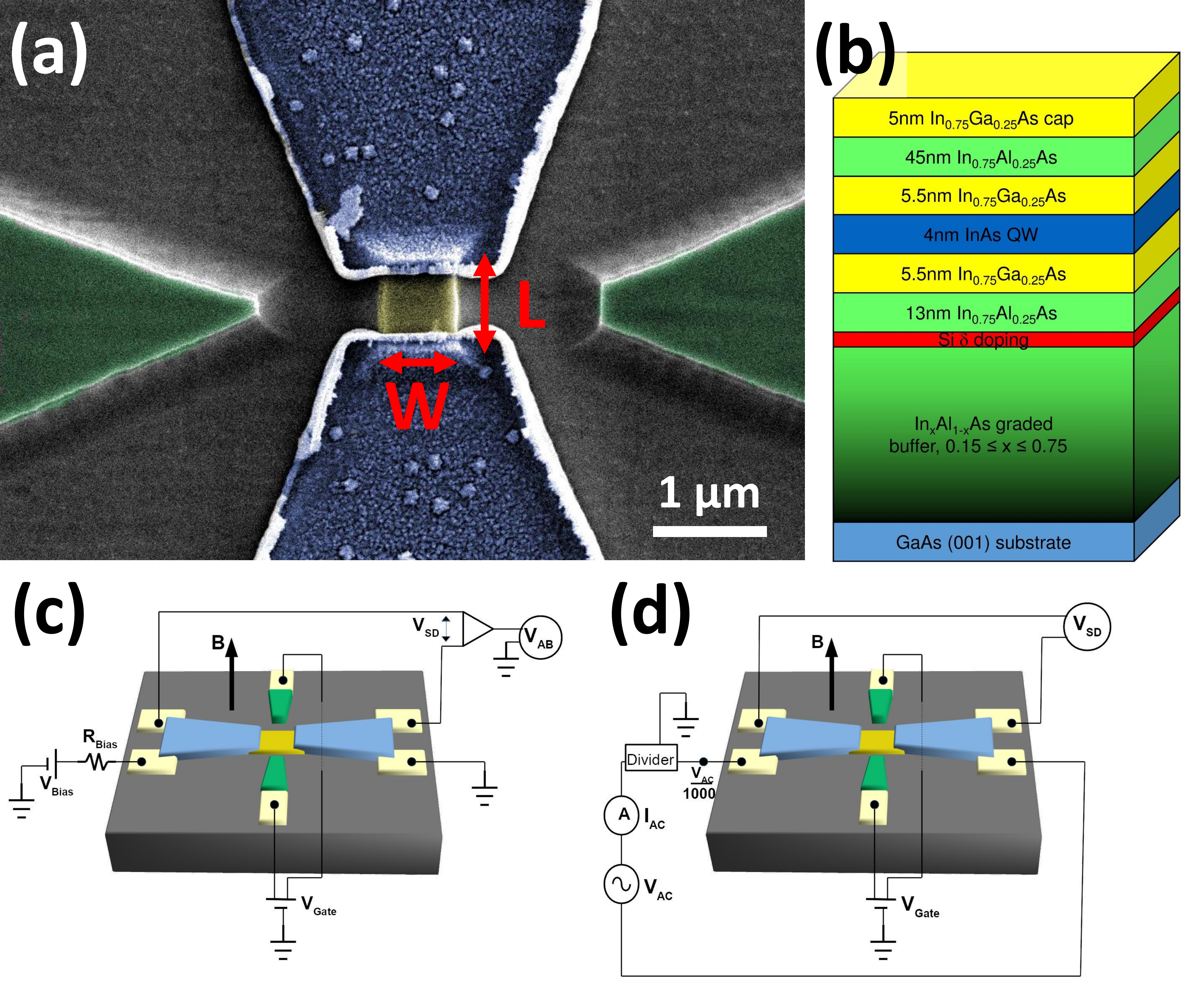}
  \caption{\label{picture:Devices1}(a) False color SEM image of a device. The mesa is yellow, side gates are green, niobium is blue. (b) Schematic drawing of the heterostructure: the 2DEG is confined in the InAs
region (blue); the Si-doped region is marked red; below there is the buffer region in order to match the InGaAs lattice constant with that of GaAs. Electrical schematics for (c) the DC current-bias and (d) the AC voltage-bias measurements. The white color indicates gold pads, the superconductor (niobium) is blue, the normal region (2DEG) is yellow, and the side gates are green. The black arrows show the direction of the magnetic field (out of plane). Filters are on each line connected to the gold pads (not shown for simplicity).}
\end{figure}

The 2DEG heterostructure was grown by Molecular Beam Epitaxy (MBE) \cite{Biasiol1,Biasiol2,Biasiol3}. As illustrated in Fig.~\ref{picture:Devices1}(b), it consists of an In$_{0.75}$Al$_{0.25}$As/ In$_{0.75}$Ga$_{0.25}$As/InAs heterostructure. The $4$~nm-thick InAs quantum well is separated from the InAlAs barriers by 5.5~nm InGaAs layers and placed $55.5$~nm underneath the surface. The n-dopant (Si) layer is $18.5$~nm below the well (modulation doping). In order to overcome the problem of lattice mismatch between In$_{0.75}$Ga$_{0.25}$As (lattice constant $a = 5.96$~\AA) and GaAs ($a = 5.65$~\AA), an InAlAs graded buffer layer was grown between the GaAs substrate and the electrically-active region. In the buffer layer, the In content is gradually increased from $15\%$ to $75\%$, substituting Al. The main parameters of the heterostructure are sheet density $n_{2D} = 6.24 \times 10^{11}$~cm$^{-2}$ (measured at $4.2$~K using Shubnikov-de Haas oscillations), mobility $\mu_e = 1.6 \times 10^5$~cm$^2$/(Vs), mean free path $l_{mfp} = 2.16$~$\mu$m, and effective electron mass $m^* = (0.030 \pm 0.001)$~m$_e$ \cite{Fornieri}.

The fabrication of our SNS devices required a sequence of mutually aligned steps of electron beam lithography (EBL) as previously reported~\cite{Fornieri,MarioSQUIPT,Mario1,Fornieri2013,Guiducci2014}. First, we defined the ohmic pads by using EBL and a standard positive-tone resist (PMMA). Then we evaporated in sequence Ni-AuGe-Ni-Au with a total thickness $\approx 200$~nm of metal. After the lift-off in acetone we annealed the contacts by heating the device up to $\approx 350$~$^\circ$C for a few minutes. Both the AuGe and the Ni layers are essential to obtain a good ohmic contact between the gold pad and the 2DEG buried below the surface. With the second lithography we defined the side gates and the mesa region of the 2DEG, i.e., the rectangular central island of the device that acts as the N region. To this end, a negative resist bilayer was spin coated on the surface of the sample and served as the mask defining the 2DEG-region and the side gates. The uncovered surface of the heterostructure was then removed by dipping in a H$_2$O:H$_2$SO$_4$:H$_2$O$_2$ solution (chemical wet etching). This procedure has also the advantage of delivering self-aligned N region and side gate in only one lithographic step. The superconducting parts of the device were fabricated with the third step of EBL. Prior to the sputter deposition of the 150~nm-thick Nb film, the interfaces were cleaned from undesired oxide layer with a dip into a HF:H$_2$O solution and a low-energy Ar$^+$ milling in the sputtering chamber itself.

Figure~\ref{picture:Devices1}(a) shows a false color scanning electron microscopy (SEM) image of a device. The yellow zone (mesa of length $L$ and width $W$) is the InAs well (2DEG). Side gates are colored in green and are formed by the same InAs layer. Blue represents niobium, i.e.~the two superconductive leads, which contact laterally the InAs quantum well. Note that part of the mesa is covered by niobium. The grey area represents the etched insulating substrate. Each niobium lead is connected to two gold pads. Two additional gold pads were formed to contact the side gates. Measurements presented here result from six devices (A -- F) that all showed a qualitatively similar behavior: their length $L$ and width $W$ are given in Table~\ref{table:LengthWidth}.

\begin{table}[t]
    \caption{\label{table:LengthWidth}Length ($L$) and width ($W$) of the measured devices.}
    \begin{ruledtabular}
		\begin{tabular}{@{}ccc@{}}
      Device & L ($\mu$m) & W ($\mu$m)  \\
      A & 0.90 & 0.73\\
      B & 0.91 & 0.65\\
      C & 1.10 & 0.85\\
      D & 0.80 & 0.90 \\
      E & 2.00 & 1.70\\
      F & 0.92 & 0.75\\    
    \end{tabular}
    \end{ruledtabular}
\end{table}

Experiments were performed at $T \approx 300$~mK in an Janis Helium-3 cryostat. Every transport line is filtered with a $\pi$-filter stage at room temperature (CMR RF-filtered breakout box) and with two stages of RC-filters positioned on the cold finger at low temperature ($T = 350$~mK), each of them with $R = 1$~k$\Omega$, $C = 47$~nF. All of them are low-pass filters, and together they have a cut-off frequency of about $f_c = 2170$~Hz. Supercurrent was measured in a 4-probe DC current-bias setup, see Fig.~\ref{picture:Devices1}(c). The bias resistor $R_{bias}$ (several M$\Omega$) maintains the source-drain current $I_{SD}$ constant, while the source-drain voltage ($V_{SD}$) is measured in 4-probe configuration and amplified with a DL preamp. On the other hand, measurements in the QH regime are performed in a 4-probe AC voltage-bias configuration, see Fig.~\ref{picture:Devices1}(d). $V_{SD}$ is measured directly with a lock-in amplifier, while the current passing through the junction, $I_{AC} = I_{SD}$, is measured with another lock-in amplifier connected in series with the voltage source. The resistance of the junction $R_{SD}$ is for both cases $R_{SD} = V_{SD} / I_{SD}$.

\section{Superconductivity}

\begin{figure}[t]
  \includegraphics[width=\linewidth]{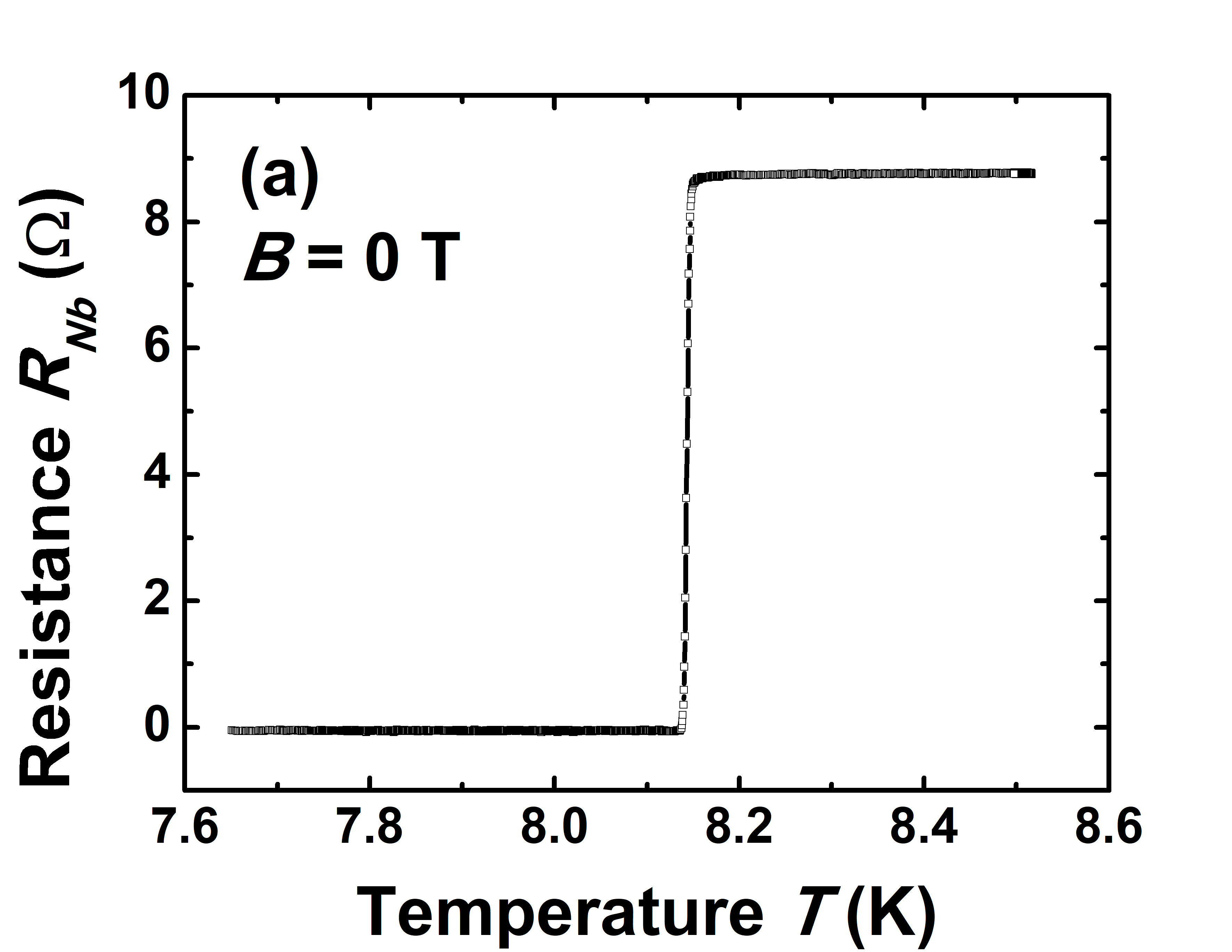}  
	\includegraphics[width=\linewidth]{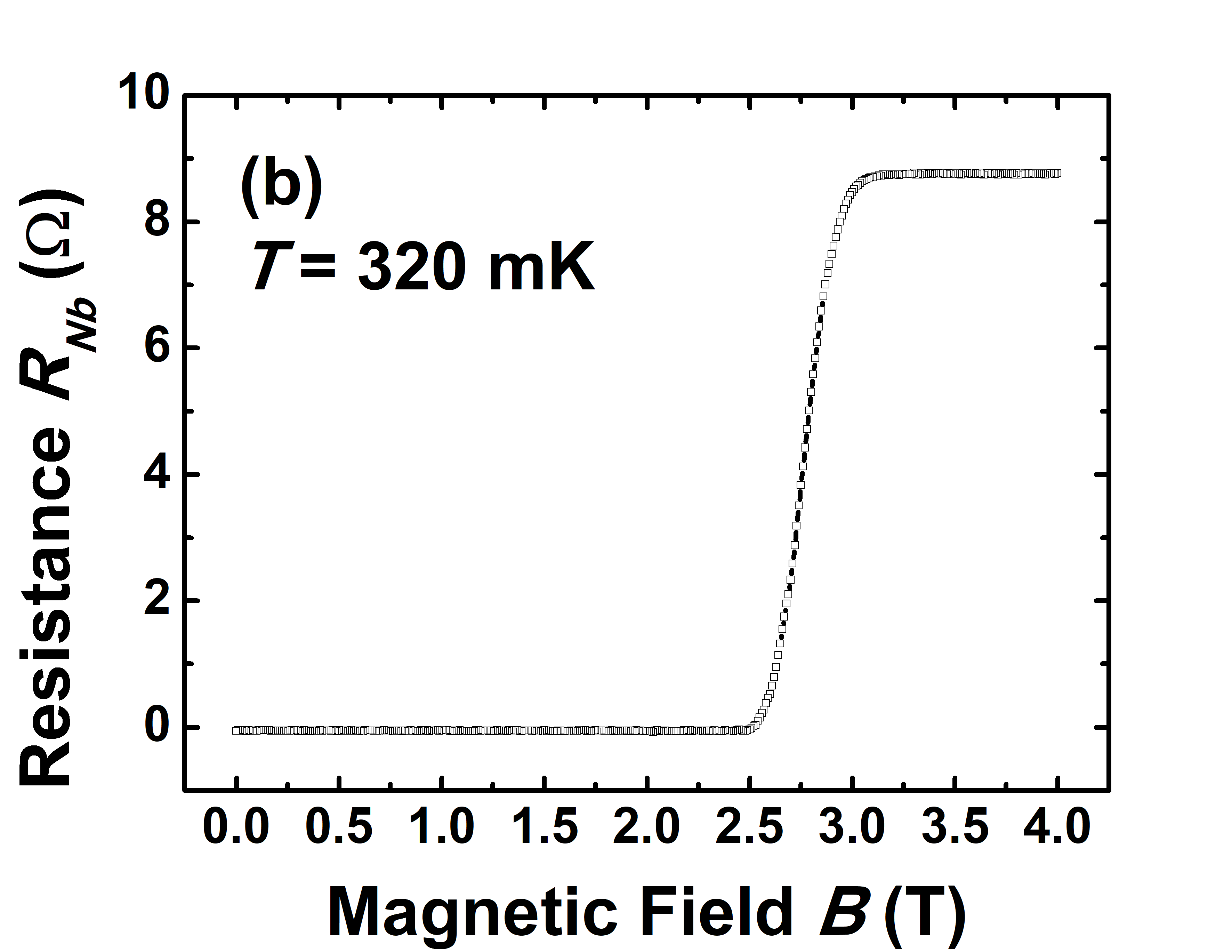}
  \caption{\label{picture:SuperNiobiumTemp}(a) Change in resistance $R_{Nb}$ as a function of temperature $T$. A sudden jump in the resistance is observed between $T=8.13$~K and $T=8.15$~K, indicating the transition from the superconductive to the normal state. The jump in resistance is about 8.8~$\Omega$. $B = 0$~T. (b) Critical field of Nb, measured at $T = 320$~mK. For both measurements, $I_{ac} = 800$~nA and $V_{gate}=0$~V.}
\end{figure}

The Niobium critical temperature $T_c$ was measured using a 4-probe AC current-bias setup on a device which actually had four gold pads connected to a Nb lead. Figure~\ref{picture:SuperNiobiumTemp}(a) shows the temperature dependence of resistance $R_{Nb}$ of the Nb. A sudden jump in resistance is observed at $T_c = (8.14 \pm 0.01)$~K, indicating the Nb transition from the superconductive to the normal state. With this value, the Nb bulk gap can be computed using BCS theory \cite{Grosso}: $\Delta_{Nb}=1.76 kT_c=1.235\pm 0.002$~meV. The resistance of the niobium leads at $8.4$~K is $R_{Nb}=(8.76 \pm 0.01)$~$\Omega$. Critical field of Nb was measured, as shown in Fig.~\ref{picture:SuperNiobiumTemp}(b). For  $T = 320$~mK, $H_c = (2.77 \pm 0.02)$~T.

The JJ devices discussed here are in the clean and ballistic regime. Indeed, using the value of the Nb gap, the coherence length of Cooper pairs in the superconductor is $\xi_0 = \frac{1}{\pi} \frac{\hbar v_F}{\Delta_{Nb}}$ \cite{Grosso}. The Fermi velocity in the superconductor is $v_F=(1.37 \pm 0.01) \times 10^6$~m/s \cite{Fornieri,Chrestin}, therefore $\xi_0 =(232 \pm 2)$~nm, and thus $\xi_0 \ll l_{mfp}$ (clean limit). Moreover, the mean free path $l_{mfp}$ of the heterostructures is larger than the length $L$ of the junctions, i.e.~$L \ll l_{mfp}$, and therefore the devices are ballistic.

\begin{figure*}[t]
  \begin{center}
  \includegraphics[width=0.49\linewidth,keepaspectratio=true]{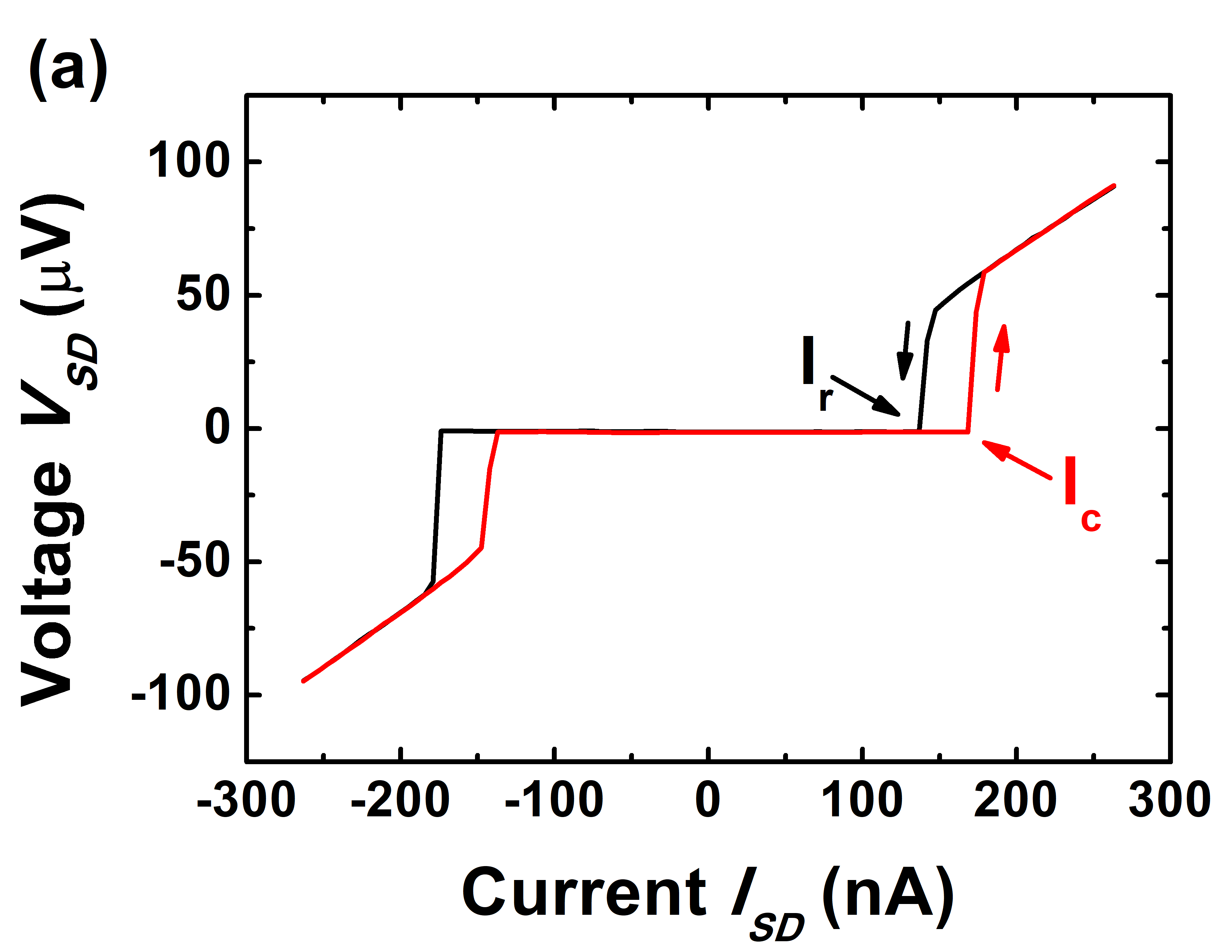}
	\includegraphics[width=0.49\linewidth,keepaspectratio=true]{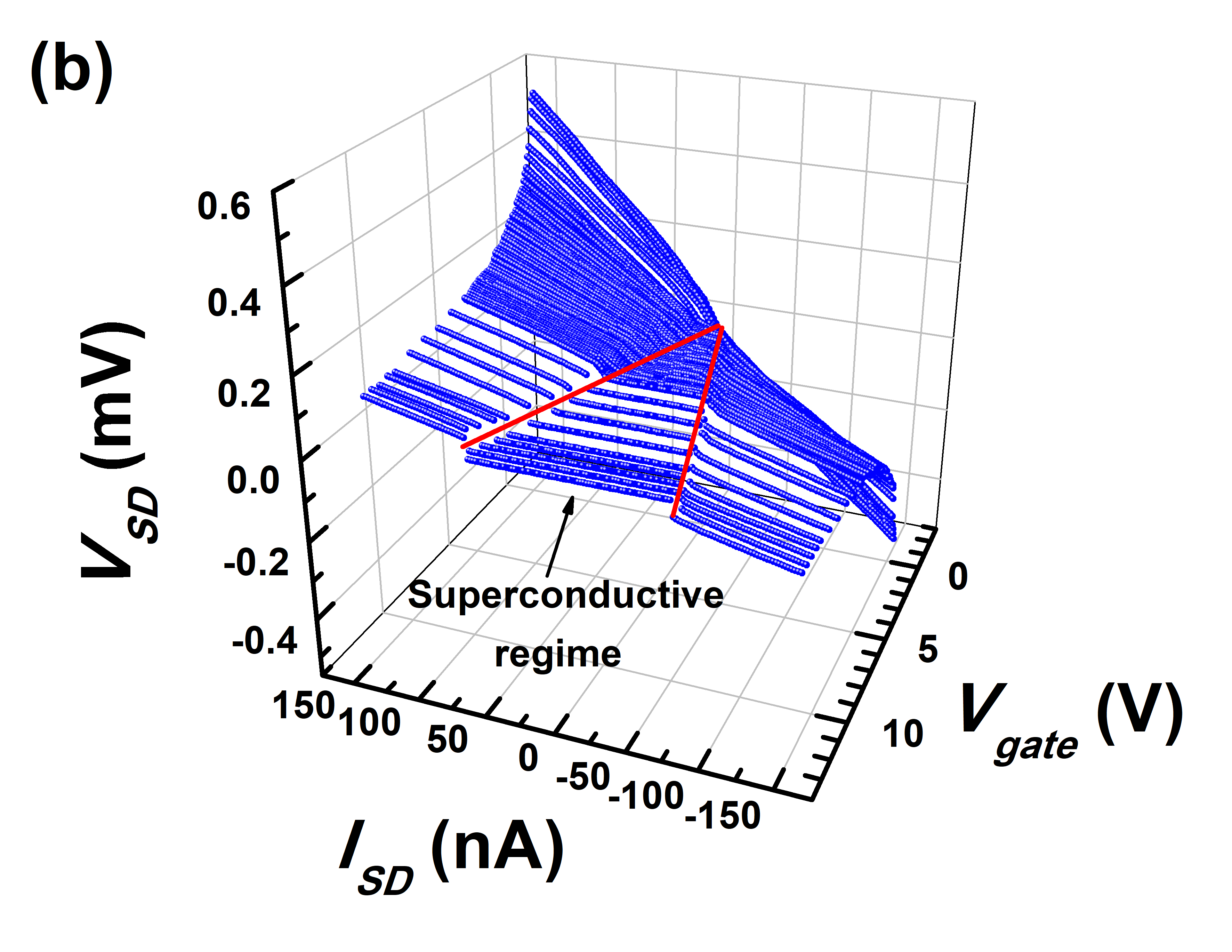}
	\end{center}
  \caption{\label{picture:SuperExample}(a) $I$--$V$ curve (source-drain voltage $V_{SD}$ vs.~source-drain current $I_{SD}$) of device C. The black (red) arrow shows the direction of the black (red) sweep. $T=400$~mK, $B = -3.5$~mT, $V_{gate}=0$~V. (b) 3D plot shows $I$--$V$ curves at different gate voltages. Two red lines indicate the transition between the superconductive and the dissipative regime. The superconductive regime becomes smaller when decreasing gate voltage; at the same time, the resistance (the slope) in the dissipative region increases with decreasing gate voltage. At about $V_{gate} = -1$~V supercurrent disappears. This measurement has been performed on device A at $B=0$~mT and $T=350$~mK.}
\end{figure*}

The JJ devices were characterized by $I$--$V$ curves, applying a DC current $I_{SD}$ through the junction and measuring the source-drain voltage $V_{SD}$ with two additional leads. All measurements were taken at the field which maximizes the supercurrent, typically a few mT, to compensate for the residual magnetization of the environment. Fig.~\ref{picture:SuperExample}(a) shows $V_{SD}$ as a function of $I_{SD}$. The flat central region with zero $V_{SD}$ represents the superconductive regime or the Josephson state. Well-developed supercurrent and typical hysteretic behavior for JJ \cite{Tinkham} are shown. From this measurement, a critical current $I_c=(170 \pm 2)$~nA and a retrapping current $I_r=(136 \pm 2)$~nA are deduced.

The normal resistance $R_n$ and the excess current $I_{exc}$ of the junctions are obtained from similar $I$--$V$ curves, but at higher injected current, in order to obtain a source-drain voltage $V_{SD} > 2 \Delta_{Nb}/e \approx 2.5$~mV. For device B, this gives $R_n = (0.75 \pm 0.01)$~k$\Omega$ and $I_{exc}=(0.57 \pm 0.01)$~$\mu$A. Similar values of $R_n$ and $I_{exc}$ were obtained for all devices. In the Octavio-Tinkham-Blonder-Klapwijk (OTBK) model \cite{OBTK,Flensberg}, they can be used to compute the height of the barrier at the normal-superconductor interface ($Z$) and the transmission probability through the junction ${\cal T}$, resulting in $Z$ between 0.81 and 0.95, and ${\cal T}$ between 52.7\% and 60.5\%. 

The two side gates yield the control of the supercurrent magnitude (both $I_c$ and $I_r$) and of the normal resistance ($R_n$) of the device \cite{Chrestin,Amado2014}. The side gates are used to increase/decrease the electron density in the 2DEG by applying positive/negative bias. Since the Josephson coupling between the electrodes and the normal resistance of the 2DEG are very sensitive to the electron density in the 2DEG (see \cite{Chrestin,Amado2014,Schapers}), the gates are expected to increase and reduce the supercurrent by increasing and decreasing $V_{gate}$, respectively. On the other hand, the normal resistance should increase (decrease) when depleting (enriching) the electron density (as seen in \cite{Chrestin,Amado2014,Schapers}). 

Figure~\ref{picture:SuperExample}(b) shows several $I$--$V$ curves at different gate voltages $V_{gate}$ in a 3D plot. The flat central region is the superconductive regime, and two red lines indicate the transition between superconductive and dissipative regime. The extension of this region is reduced by decreasing $V_{gate}$ (supercurrent reduced), until the supercurrent vanishes below $V_{gate} = -1$~V. At the same time, the slope (i.e. the resistance) in the dissipative region increases when decreasing $V_{gate}$.

These data show that the magnitude of the supercurrent can be controlled by the side gates, forming a so-called Josephson Field Effect Transistor (JoFET) \cite{Akazaki1996,Clark1980,Bezuglyi2017}. A similar reduction of supercurrent is expected upon increase in temperature. Temperature-dependent measurements clearly show that, as temperature is increased, the superconductive region becomes smaller. For device D, the supercurrent vanishes above $T=650$~mK. At higher temperatures, a slope different from zero appears even at the origin, i.e. the resistance is different from zero ($R(I_{SD} \approx 0$~nA)~$\neq0$~$\Omega$); however, there is still a kink on both sides of the origin. At temperatures even higher (4 K for device D), also this kink disappears, and the $I$--$V$ becomes completely linear (i.e. Ohmic). A similar behaviour was found in  all devices.

\section{Quantum Hall Regime}

\begin{figure*}[t]
  \includegraphics[width=0.49\linewidth,keepaspectratio=true]{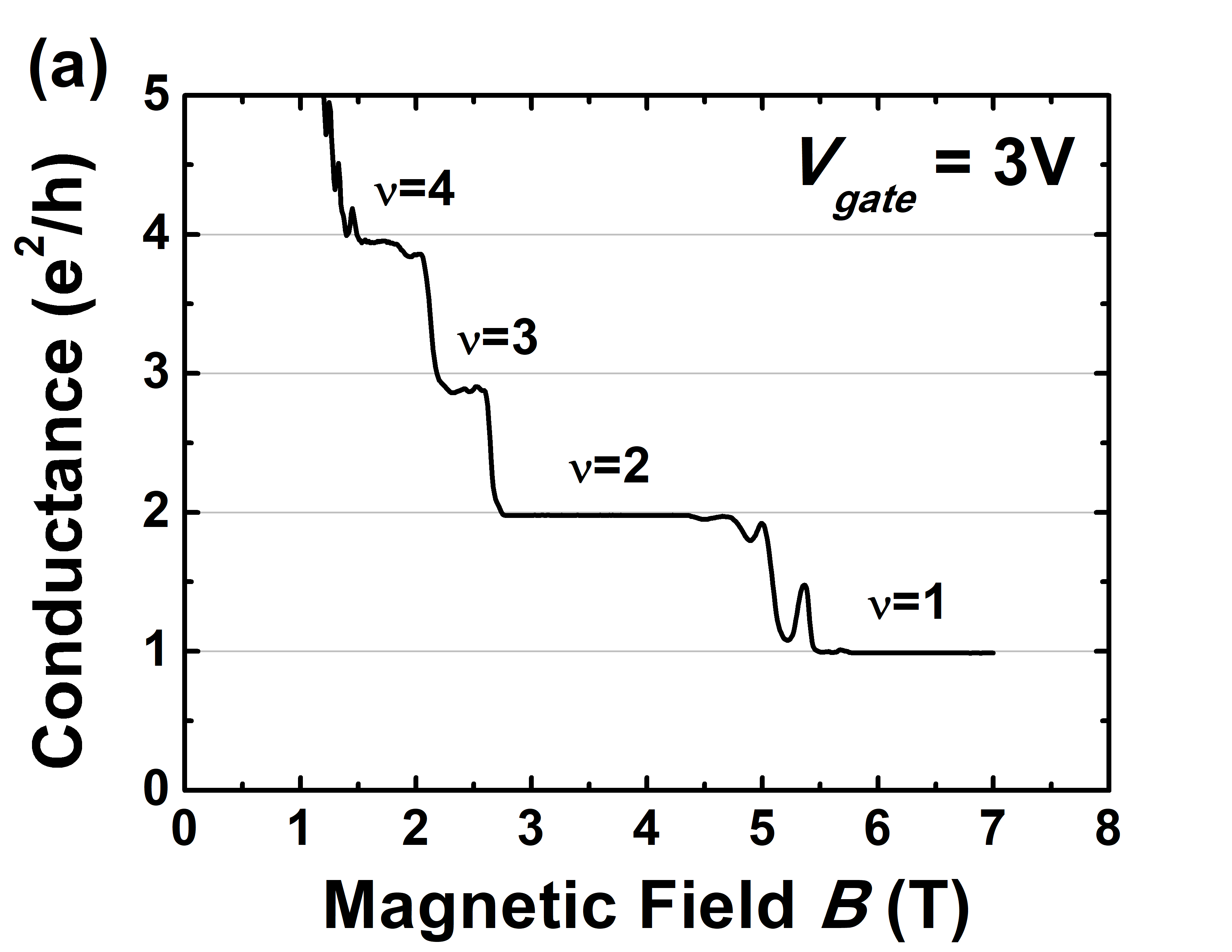}
	\includegraphics[width=0.49\linewidth,keepaspectratio=true]{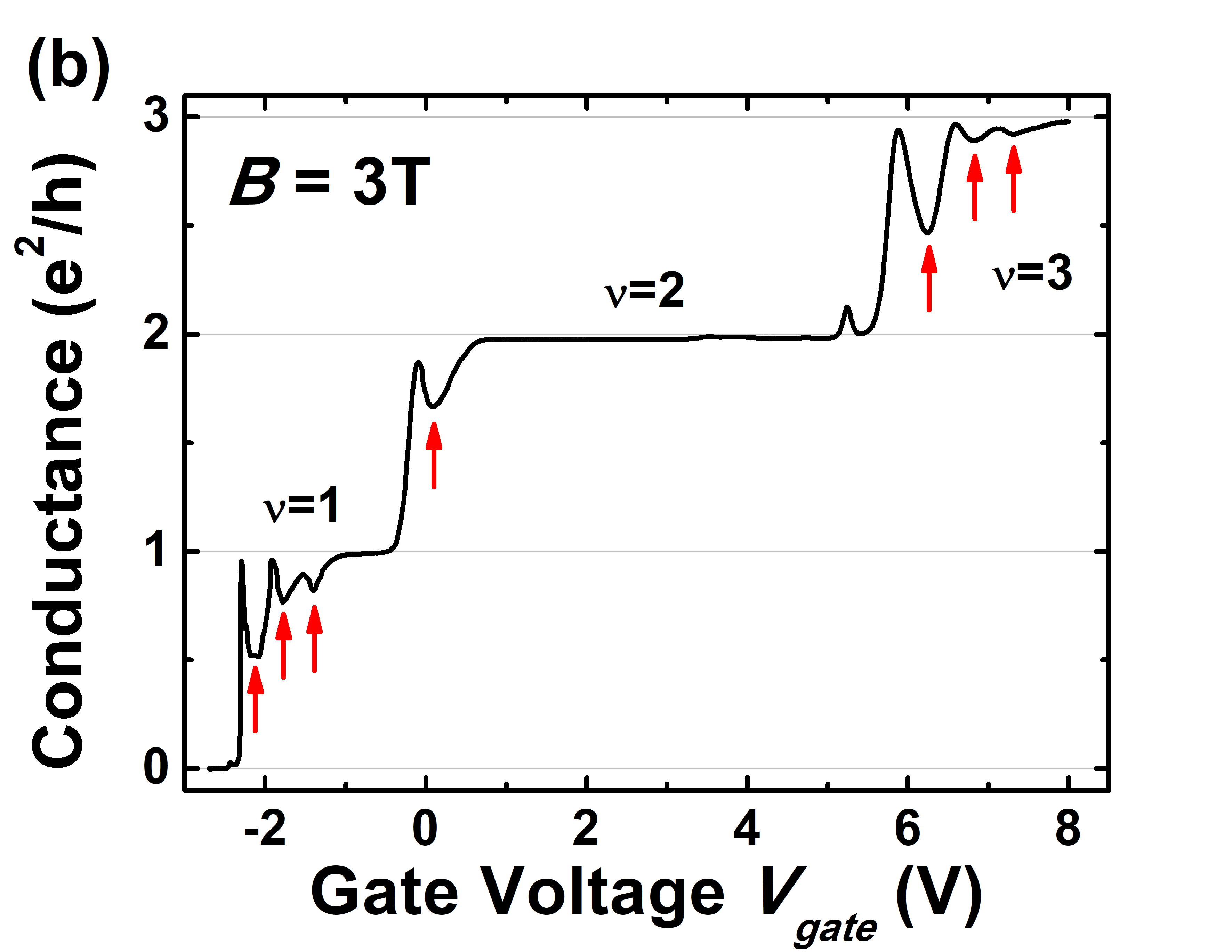}
  \caption{\label{picture:SGMgate1}(a) Conductance (in units of $e^2/h$) as a function of magnetic field, performed on device B at $V_{gate} = 3$~V. (b) Conductance as a function of $V_{gate}$, performed at $B = 3$~T. Resonances are highlighted with red arrows. $T=433$~mK.}
\end{figure*}

Figure~\ref{picture:SGMgate1}(a) reports the conductance of a representative device as a function of magnetic field, clearly showing four quantum Hall plateaus. As the magnetic field is increased, the degeneracy of Landau levels increases, therefore less levels are occupied at higher field, i.e.~conductance decreases. The quantum Hall regime is established at $B = 1.5$~T, i.e.~this is the smallest field at which plateaus are well-developed. Notice that at $B = 3$~T and $V_{gate} = 3$~V, the filling factor is $\nu = 2$.

Next, the magnetic field was fixed at $B = 3$~T, and the gate voltage $V_{gate}$ was swept until the mesa was pinched off. Figure~\ref{picture:SGMgate1}(b) shows conductance as a function of $V_{gate}$. At $V_{gate} = 8$~V, $\nu=3$ is measured. Since more Landau levels are occupied with respect to $V_{gate} = 3$~V, side gates have globally increased electron density in the 2DEG. Therefore, the side gates control both the width of the constriction and the electron density in the whole 2DEG. As $V_{gate}$ is decreased, Landau levels are progressively emptied, so $\nu$ decreases (in a step-like fashion) until conductance is zero at $V_{gate} = -2.6$~V. 

Interestingly, Fig.~\ref{picture:SGMgate1}(b) shows that in particular for $\nu=1$ and $\nu=3$, several resonances are well-developed at the end of plateaus. These features can be explained by properly modeling the shape of the electrostatic potential generated by side gates \cite{Guiducci2014}. Physically, an abrupt and steep profile of the electrostatic potential leads to the interference of edge states thus reducing the transmission probability of the system for certain energy values of the barrier in a Fabry-P\'erot-like behavior. The observed behavior is in good agreement with simulations performed on similar systems \cite{Palacios}.

\begin{table}[b]
  \caption{\label{tab:densityVsGate}Electron density $n_s$ vs.~gate voltage $V_{gate}$ performed on device E at $T = 350$~mK.}
	\begin{ruledtabular}
  \begin{tabular}{@{}cc@{}}
    $V_{gate}$ ($V$) & $n_s$ ($ 10^{11} cm^{-2}$)  \\
    $+4$ & $1.94 \pm 0.08$ \\
    $0$ &  $1.57 \pm 0.07$ \\
    $-1$ & $1.45 \pm 0.07$  \\
    $-1.7$ & $1.05 \pm 0.06$  \\
  \end{tabular}
  \end{ruledtabular}
\end{table}

These data can be used to obtain the carrier density in the quantum well. Since on the plateaus the longitudinal resistance vanishes, there the source-drain resistance equals the transverse Hall resistance, $R_{SD} = R_H =  \frac{B}{n_s e}$ and can be used to compute the electron density $n_s$ in the 2DEG. For device B, $n_s = (1.82 \pm 0.04) \times 10^{11}$~cm$^{-2}$, indicating a slight reduction in carrier density with respect to the pristine heterostructure. The same measurement was also performed on device E, but this time several magnetic field sweeps were taken at different $V_{gate}$ in order to see if (and how) carrier density changed with gate voltage (see Table~\ref{tab:densityVsGate}). Data show that carrier density does decrease when decreasing gate voltage. In particular, depletion is highly non-linear and drops below $V_{gate} = -1$~V. This clearly shows that the side gates can effectively control electron density of the 2DEG inside the InAs quantum well.

\section{Conclusions}
Fabrication and analysis of JJ devices within a high-mobility InAs quantum well and Nb superconducting contacts have been reported. Transport measurements demonstrated Josephson current flowing in these structures that also remarkably exhibit a critical temperature $T_c\sim 8.1$~K and field $H_c\sim 3$~T. Side gates nearby the JJ allow for the tuning of superconducting properties, i.e. modulation of critical current $I_c$ as a function of gate voltage. We have shown that the same devices present well-developed QH plateaus, measured at $T\sim 300$~mK, reaching filling factor $\nu=2$ at $B=3$~T. Gate voltage response in the QH regime has also been  reported, demonstrating control of channel width and electronic density. We believe this study provides a useful platform for the investigation of hybrid systems where the coexistence of superconductivity and QH features can lead to the formation of topological states, such as Majorana bound states. Further material optimization can lead to the study of the fractional QH regime in proximity to SC contacts, allowing for the investigation of exotic states like parafermions \cite{Clarke2013,Lindner2012,Clarke2014}.

\begin{acknowledgments}
We are indebted to Francesco Giazotto for his support in the initial phase of this research. S.~G.~acknowledges support by Fondazione Silvio Tronchetti Provera. M.~ C.~ and L.~ S.~ acknowledge support from the Quant-Eranet project "SuperTop". S.~H.~acknowledges support from Scuola Normale Superiore, project SNS16\_B\_HEUN---004155. We acknowledge funding from the Italian Ministry of Foreign Affairs, Direzione Generale per la Promozione del Sistema Paese (agreement on scientific collaboration between Italy and Poland). Financial support from the CNR in the framework of the agreement on scientific collaboration between CNR and CNRS (France) is acknowledged. Furthermore, this work was financially supported by EC through the project PHOSFUN {\it Phosphorene functionalization: a new platform for advanced multifunctional materials} (ERC ADVANCED GRANT No.~670173). 
\end{acknowledgments}

%
\bibliographystyle{pss}
\bibliography{Bibliography}

\end{document}